\documentclass{pasj01}
\Received{$\langle$2016/12/22$\rangle$}
\Accepted{$\langle$2017/02/dd$\rangle$}
\Published{$\langle$yyyy/mm/dd$\rangle$}
\usepackage{color}
\usepackage{lscape}
\usepackage{threeparttable}
\newcommand{\cii}{C{\sc ii}}

\begin{document}
\title{ALMA Deep Field in SSA22: Blindly Detected CO Emitters and [\cii] Emitter Candidates}
\author{
N.~H.~Hayatsu\altaffilmark{1*}, 
Y.~Matsuda\altaffilmark{2, 3}, 
H.~Umehata\altaffilmark{4, 5}, 
N.~Yoshida\altaffilmark{1, 6}, 
I.~Smail\altaffilmark{7}, 
A.~M.~Swinbank\altaffilmark{7}, 
R.~Ivison\altaffilmark{8}, 
K.~Kohno\altaffilmark{5, 9}, 
Y.~Tamura\altaffilmark{5}, 
M.~Kubo\altaffilmark{10}, 
D.~Iono\altaffilmark{2, 3}, 
B.~Hatsukade\altaffilmark{2}, 
K.~Nakanishi\altaffilmark{2, 3},
R.~Kawabe\altaffilmark{2, 3}, 
T.~Nagao\altaffilmark{11}, 
A.~K.~Inoue\altaffilmark{12}, 
T.~T.~Takeuchi\altaffilmark{13}, 
M.~Lee\altaffilmark{2, 14},
Y.~Ao\altaffilmark{2},
S.~Fujimoto\altaffilmark{14, 15},
T.~Izumi\altaffilmark{5}, 
Y.~Yamaguchi\altaffilmark{5}, 
S.~Ikarashi\altaffilmark{16}, 
and
T.~Yamada\altaffilmark{17, 18}
}

\altaffiltext{}{
\altaffilmark{1} Department of Physics, The University of Tokyo, 7-3-1 Hongo, Bunkyo, 
			Tokyo 113-0033, Japan\\
\altaffilmark{2} National Astronomical Observatory of Japan, Osawa 2-21-1, 
			Mitaka, Tokyo 181-8588, Japan\\
\altaffilmark{3} Graduate University for Advanced Studies (SOKENDAI), 
			Osawa 2-21-1, Mitaka, Tokyo 181-8588, Japan\\
\altaffilmark{4}	The Open University of Japan, 2-11 Wakaba, Mihama-ku, Chiba 261-8586, Japan\\
\altaffilmark{5} Institute of Astronomy, School of Science, The University of Tokyo, 2-21-1 
			Osawa, Mitaka, Tokyo 181-0015, Japan\\
\altaffilmark{6} Kavli Institute for the Physics and Mathematics of the Universe (WPI),
			Todai Institutes for Advanced Study, \\
			The University of Tokyo, Kashiwa, Chiba 277-8583, Japan\\
\altaffilmark{7} Centre for Extragalactic Astronomy, Department of Physics, Durham University, 
			South Road, Durham, DH1 3LE, UK\\
\altaffilmark{8} European Southern Observatory, Karl-Schwarzschild-Str. 2, D-85748 Garching, Germany\\
\altaffilmark{9} Research Center for the Early Universe, The University of Tokyo, 7-3-1 
			Hongo, Bunkyo, Tokyo 113-0033\\
\altaffilmark{10} National Astronomical Observatory of Japan TMT Project Office, 
			Osawa 2-21-1, Mitaka, Tokyo 181-8588, Japan\\
\altaffilmark{11} Research Center for Space and Cosmic Evolution, Ehime University, 
		Matsuyama 790-8577, Japan\\
\altaffilmark{12} College of General Education, Osaka Sangyo University, 3-1-1 Nakagaito, Daito, 
			Osaka 574-8530, Japan\\
\altaffilmark{13} Division of Particle and Astrophysical Science, Nagoya University, 
			Furo-cho, Chikusa-ku, Nagoya 464-8602, Japan\\
\altaffilmark{14} Department of Astronomy, Graduate school of Science, The University of Tokyo, 
			7-3-1 Hongo, Bunkyo-ku, Tokyo 133-0033, Japan\\
\altaffilmark{15} Institute for Cosmic Ray Research, University of Tokyo, 
			5-1-5 Kashiwa-no-Ha, Kashiwa City, Chiba 277-8582, Japan\\
\altaffilmark{16} Kapteyn Astronomical Institute, University of Groningen, P.O. Box 800, 
			9700AV Groningen, The Netherlands\\
\altaffilmark{17} Astronomical Institute, Tohoku University, 6-3 Aoba, Aramaki, 
			Aoba-ku, Sendai, Miyagi 980-8578, Japan\\
\altaffilmark{18} Institute of Space and Astronautical Science, JAXA, 3-1-1 Yoshinodai, Sagamihara, Kanagawa Japan
}
\email{natsuki.hayatsu@utap.phys.s.u-tokyo.ac.jp}
\maketitle

\begin{abstract}
We report the identification of four millimeter line emitting galaxies 
with the Atacama Large Milli/submillimeter Array (ALMA) in SSA22 Field (ADF22).  
We analyze the ALMA 1.1 mm survey data, with an effective survey area 
of 5 arcmin$^2$, a frequency range of 253.1--256.8 and 269.1--272.8 GHz, 
angular resolution of 0$''$.7 and RMS noise of 0.8 mJy beam$^{-1}$ at 36 km s$^{-1}$ velocity resolution.  
We detect four line emitter candidates with significance levels above $6 \sigma$.  
We identify one of the four sources as a CO(9-8) emitter at $z = 3.1$ in a member of the proto-cluster known in this field.  
Another line emitter with an optical counterpart is likely a CO(4-3) emitter at $z = 0.7$.  
The other two sources without any millimeter continuum or optical/near-infrared counterpart 
are likely to be [\cii] emitter candidates at $z = 6.0$ and $6.5$. 
The equivalent widths of the [\cii ] candidates are consistent with those of confirmed high-redshift 
[\cii] emitters and candidates, 
and are a factor of 10 times larger than that of the CO(9-8) emitter detected in this search.  
The [\cii] luminosity of the candidates are $4-7 \times 10^8~\rm L_\odot$.  
The star formation rates (SFRs) of these sources are estimated to be 
$10-20~\rm M_{\odot}~yr^{-1}$ if we adopt an empirical [\cii] luminosity - SFR relation.  
One of them has a relatively low-S/N ratio, but shows features characteristic of emission lines. 
Assuming that at least one of the two candidates is a [\cii] emitter, we derive a lower limit of [\cii]-based 
star formation rate density (SFRD) at $z~\sim~6$. The resulting value of $> 10^{-2}$ 
$\rm M_\odot yr^{-1}  Mpc^{-3}$ is consistent with the dust-uncorrected UV-based SFRD.  
Future millimeter/submillimeter surveys can be used to detect a number of high redshift line emitters, with which 
to study the star formation history in the early Universe.  
\end{abstract}

\section{Introduction}
	The cosmic star-formation history in the early Universe 
	has been studied in optical/near-infrared (NIR) wavelengths, 
	which trace ultraviolet (UV) radiation in rest-frame at high redshifts (e.g.,\,\cite{madau2014}).  
	The UV star formation rate density (SFRD) does not account for all components of 
	star-forming galaxies (e.g.,\,\cite{bouwens2012,bouwens2016}).
        Recent studies suggest that far-infrared (FIR) SFRD
	contributes more than half of the total at $z = 0-4$ 
	(e.g.,\,\cite{blain1999, barger2012, burgarella2013, gruppioni2013, swinbank2014}). 
	Millimeter/submillimeter (mm/submm) galaxy surveys would be, in principle,
	efficient to probe the dust-obscured component of SFRD at high-redshift 
	\citep{takeuchi2005, burgarella2013, chen2016, carniani2015, fujimoto2016, 
	aravena2016a, dunlop2017,umehata2017}.  
	The advantage of such observations in mm/submm is the well-known negative $k$-correction; 
	the continuum flux of a typical star-forming galaxy of fixed SFR remains 
	approximately constant with increasing redshift \citep{blain2002}.  
	However, it is often difficult to estimate redshifts for very faint and dusty sources 
	(e.g.,\,\cite{simpson2014}). 

	Strong emission lines such as [\cii]158$\rm\mu$m or [O{\sc iii}] 88$\rm\mu$m lines can be used to study 
	the SFR and gas properties of high-$z$ star-forming galaxies as well as to determine their 
	spectroscopic redshifts
	(e.g.,\,\cite{colbert1999, maiolino2005, brauher2008, swinbank2012, venemans2012, delooze2014, 
	inoue2014, willot2015, maiolino2015, inoue2016, carniani2017}).  
	Interestingly, \citet{capak2015} report that Lyman-break galaxies (LBGs) at $z = 5-6$ 
	show enhancement of [\cii] emission relative to the FIR continuum 
	compared with mm/submm-selected galaxies.  
	They also serendipitously detected a [\cii] emitter which is faint in both the 
	rest-UV and FIR continuum.  Combining observations in rest-UV, FIR and mm/submm emission lines 
	appears to be essential to understand the physical properties
         of galaxies at high redshifts (e.g.,\,\cite{bouwens2016, aravena2016a, dunlop2017}). 
	
	One of the brightest submm emission lines is [\cii] 
	(e.g.,\,\cite{maiolino2005, maiolino2009, iono2006, venemans2012, 
	swinbank2012, willott2013, willot2015, maiolino2015, capak2015, diaz2016, pentericci2016}).  
	Carbon in the interstellar medium is largely in a singly ionised state in a variety
        of environments, from H{\sc ii} regions to molecular clouds, 
	because the ionization potential of atomic carbon is 11.3 eV, lower than that of hydrogen 
	and close to dissociation energy of CO of 11.1 eV (e.g.,\,\cite{wolfire2010, carilli2013}).  
	The critical density of [\cii] emission is about $3\times10^3$ cm$^{-3}$, 
	and thus [\cii] emission can arise even in a molecular cloud 
	with temperature around 92 K \citep{hollenbach1989}.  
	Therefore [\cii] radiative cooling often dominates in regions with a wide range of
        densities (e.g.,\,\cite{wolfire1995, kaufman1999}).  
	Finally, [\cii] emission is thought to be a potential tracer of SFR 
	because of its main origin of photo-dissociated region associated with young,
        massive stars (e.g.,\,\cite{delooze2011, delooze2014, sargsyan2012, kapala2015}).  
	An important observational advantage is that [\cii] line emission at $z >  4$ is
        redshifted to wavelengths with
	low atmospheric absorption and thus it is possible to detect [\cii] line
	emission even from galaxies at $z = 7$ (e.g.,\,\cite{venemans2012,aravena2016b,pentericci2016}).  

        A number of high-redshift [\cii] 
	emitters are expected to be detected with forthcoming high sensitivity observations 
	with the Atacama Large Millimeter/submillimeter Array (ALMA) 
	(e.g.,\,\cite{geach2012, cunha2013, matsuda2015, aravena2016b}).  
	In this paper, we present a blind search for [\cii] emitters 
	using ALMA Cycle 2 data \citep{umehata2017}. 
	We briefly introduce the observations in \S 2. The details of our data analysis
        is described in \S 3.  
	Then we show the results in \S 4 and discuss the implications for cosmic star formation history in \S 5.  
	We summarize the results and discussions in \S 6.  
	Throughout the paper, we adopt the standard
        $\rm \Lambda$CDM cosmology with the matter density 
	$\Omega_{\rm M} = 0.3$, the cosmological constant $\Omega_{\rm \Lambda} = 0.7$, 
	the Hubble constant $h = 0.7$ in the unit of $H_0 = 100~\rm km~s^{-1} Mpc^{-1}$.  
	All magnitudes are given in the AB system, unless otherwise noted.  
	We calculate SFR assuming Chabrier initial mass
        function (IMF) \citep{chabrier2003}, 
        with an integration range from 0.08 $\rm M_\odot$ to 100 $\rm M_\odot$.  
	When needed, we use the conversion factor of 1.8 from the Chabrier IMF 
	to the equivalent Salpeter IMF \citep{salpeter1955} 
	and 1.1 from the Chabrier IMF to the Kroupa IMF \citep{kroupa2001}.  

\section[]{Observation}
\begin{figure*}
\begin{center}
\includegraphics[trim=30 80 60 60, width=120mm]{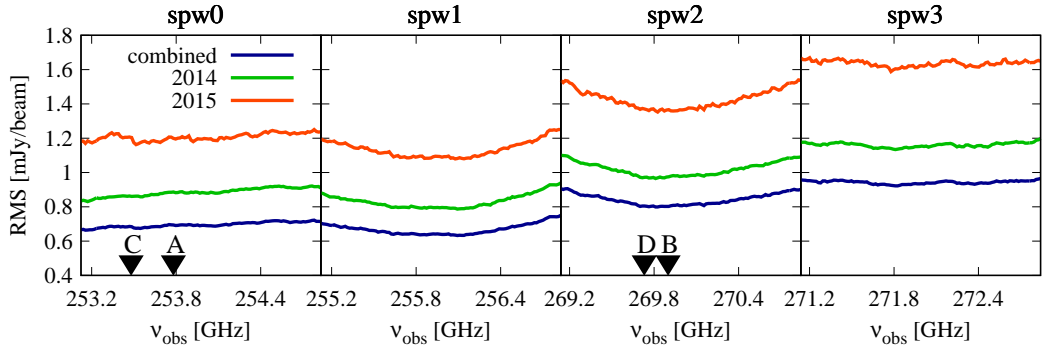}
\end{center}
  \caption{
   The RMS noise level of the four spectral windows (SPWs) analyzed in this work 
  at 36 km s$^{-1}$ velocity resolution 
  as a function of observed frequency.  
  In each panel, we plot the RMS of combined data and the individual 
  data obtained in 2014, and in 2015, respectively.  
  We also mark the frequencies of the detected four line emitter candidates A--D.  
   Typical RMS values are also given in Table 1. We note that the data observed 
   in 2014 and 2015 have different angular resolutions.  
}
\end{figure*}

	We analyze data from the ALMA Deep Field survey of SSA22 (ADF22) observed 
	in Band 6 in ALMA Cycle 2 in June 2014 and April 2015 
	(Proposal ID 2013.1.00162.S, PI: H. Umehata).  
	The details of the observation are described in \citet{umehata2017}.  
	
	ADF22 is a survey field with an area of $2' \times 3'$ centered on 
	a $z = 3.09$ proto-cluster; RA (J2000) = 22$\rm ^h$17$\rm ^m$34$\rm^ s$,  
	Dec (J2000) = +00$^\circ$17$'$00$''$ consisting of 103 pointing fields.  
	The field was observed using four 1.875 GHz spectral windows (SPW) 
	with the central frequency of 263 GHz, 
	which corresponds to the [\cii] redshift of $z = 6.2$.  
	
	The typical angular resolution of a combined data is $0''.72 \times 0''.62$ corresponding
        to $\sim 6$ kpc at $z = 6.2$.  
        The on-source time per pointing in the fields is 4.5 min. 
        The data observed in 2014 and 2015 have angular resolution of 
   	$0''.54 \times 0''.49$ and $1''.24 \times 0''.87$, 
	and on-source time per pointing in the fields of 2.5 min. and 2.0 min.
	for 2014 and 2015, respectively. 

	The four SPWs have root-mean-square (RMS) noise level of 
	0.7, 0.7, 0.8 or 0.9 mJy beam$^{-1}$ at a 36 km s$^{-1}$ velocity resolution.  
	The RMS of each SPW at 36 km s$^{-1}$ resolution of combined, 2014 and 2015 data
	as a function of the observed frequency 
	are shown in Figure 1, where no significant atmosphere absorption is seen.  
	Other properties of the data are listed in Table 1.  

	In order to search faint emission line sources, we use high sensitivity data 
	of 80 pointing fields; Field 1 - Field 80 and search in a rectangle
        area of $\sim$ 5 arcmin$^2$; 
	(RA (J2000), Dec (J2000)) = (22$\rm ^h$17$\rm ^m$31.86$\rm^ s$, +00$^\circ$15$'$25.46$''$) 
	to (22$\rm ^h$17$\rm ^m$38.17$\rm^ s$, +00$^\circ$18$'$35.05$''$), 
	and a frequency coverage of 253.1--272.8 GHz (Table 1). 
	The effective survey area corresponds to about 29 comoving Mpc$^2$ 
	and the effective survey volume is $\sim$ 2.2 $\times$ $10^3$ comoving Mpc$^3$ at $z = 6.2$.  

\section[]{Method}
	The flowchart of our source selection method is shown in Figure 2.  
	The data are analyzed with Common Astronomy Software Application 
	({\sc casa}) ver.\,4.5.3 \citep{mcmullin2007}.  
	We make continuum-subtracted datacube by using {\sc uvcontsub} and {\sc clean}.  
	We first spectrally smooth the data to obtain high signal-to-noise (S/N) ratios.  
	The top-hat spectral smoothing window is set to be 0, 2, 4, ..., 12, 15, 18, ..., 21 slices, 
	with a slice width corresponding to $\sim$18 km s$^{-1}$.  
	We use the spectral smoothing function ``boxcar'' so that 
	the velocity sampling of the output data is kept constant.  
	As each spectral data slice has a different RMS value as shown in Figure 1,  
	we normalise each slice by its RMS.  
	We call a datacube thus-generated as ``S/N cube''.  

	We use {\sc clumpfind} \citep{williams1994} to search emission line sources in the S/N cube.  
	We search for sources with a threshold value ``low'' of {\sc clumpfind} of $\geq$ 4.5.  
	We then do `matching' of the clumps detected at the same position 
	between the S/N cubes in the same SPW with different resolutions
 	and retain the clump that has maximum S/N ratio (see also Table 1).  
	We select clumps that have the S/N ratio larger than $6.0 \sigma$ and also
        larger than the maximum negative S/N ratio measured
        in the inverted S/N cube in each SPW (see also Figure 2), 
	in order to avoid contamination by spurious sources (e.g.,\,\cite{hatsukade2016}).  
	We also check line spectral features of the detected clumps 
	(sources) in the datacube separately for those observed in 2014 and 2015.  

	For the detected sources, we search for their counterparts in 
	$u^*$ band taken with the Canada France Hawaii Telescope/MegaCam obtained by archival data 
	\citep{kousai2011}, $B$, $V$, $R$, $i'$, $z'$, NB912, 
	$J$, $H$, $K$ band taken with the Subaru Telescope \citep{hayashino2004,nakamura2011,suzuki2008,uchimoto2012}, 
	3.6 $\rm \mu$m, 4.5 $\rm \mu$m, 5.8 $\rm \mu$m, 8.0 $\rm \mu$m, 
	24  $\rm \mu$m taken with the {\it Spitzer} Space Telescope/IRAC and MIPS \citep{hainline2009,webb2009}
	0.5 keV, 2 keV and 8 keV taken with the {\it Chandra} X-Ray Observatory \citep{lehmer2009}.  
	
	[\cii] line emitting galaxies at $z = 6.0-6.5$ are likely to be 
	detected only longward of $z'$ band and/or in narrow-band NB912 if they are LBGs or 
	Ly$\rm \alpha$ emitters (LAEs) (e.g.,\,\cite{nakamura2011}), 
	although the available $z'$ band and NB912 data could be too shallow for
	high-redshift [\cii] emitters in our blind search.  
	For the sources with counterparts, we estimate either their photometric redshift 
	by means of spectral energy distribution (SED) fitting or spectroscopic redshift 
	by assuming their line species. 
	SED fitting is calculated by using {\sc hyperz} software \citep{bolzonella2000}.  
	In \S 4.3, we also use the equivalent width and the source number density
	to consider if the detected [\cii] emitter candidates are other line emitters.  
\begin{figure}[tbh]
	\begin{center}
	\includegraphics[trim=0 0 0 0, width=85mm]{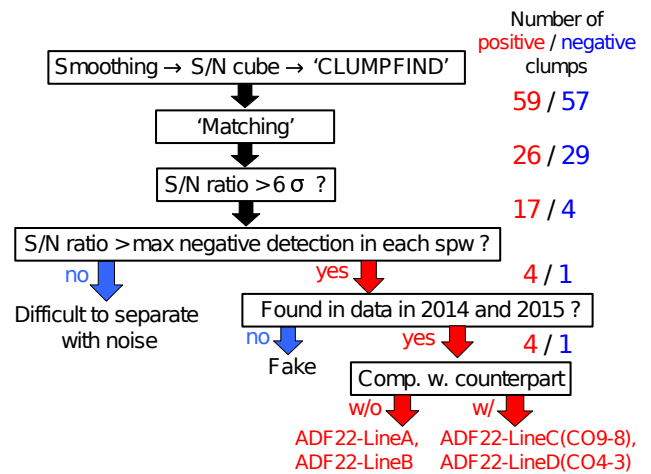}
	\end{center}
	  \caption{
	  Flowchart of the our selection method.  
	  The number of the retained clumps with $> 5.5 \sigma$ at each step is given on the right.  
	  We select the targets by setting a S/N threshold $6.0 \sigma$ in each SPW, 
	  and then by imposing that their S/N are larger than the maximum negative
          S/N ratio (see also Table 1).  
	  Finally, the selection leaves four clumps as line emitting galaxy candidates. 
	  }
\end{figure}

\section[]{Result}
\subsection[]{Source Detection}
\begin{figure*}
	\begin{center}
	\includegraphics[trim=70 -20 100 10, width=180mm]{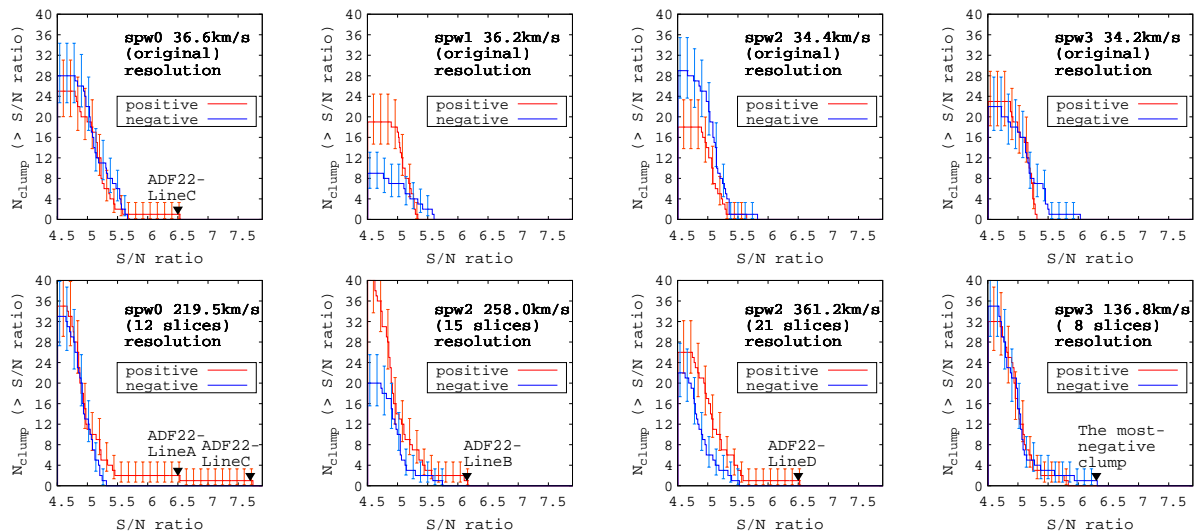}
	\end{center}
	\caption{
	Cumulative number of the positive and negative clumps as a function of S/N, 
	with $1\sigma$ error bar from the source number statistics \citep{gehrels1986}.  
	We use the continuum-subtracted S/N cubes before `matching'.  
	Top panels show the number distributions with our fiducial spectral resolution. 
	ADF22-LineC is detected with $6.5 \sigma$ in S/N cube at the fiducial resolution.  
	Bottom panels show the result with smoothed spectral resolutions. 
	ADF22-LineA and ADF22-LineC are detected in the S/N cube 
	at 220 km s$^{-1}$ spectral smoothing, 
	ADF22-LineB at 258 km s$^{-1}$ spectral smoothing, 
	and ADF22-LineD at 361 km s$^{-1}$ spectral smoothing.  
	ADF22-LineD is also detected with $6.1 \sigma$ at 258 km s$^{-1}$ spectral smoothing.  
	The most-negative clump detected in SPW 3 has a S/N ratio of $6.3 \sigma$ in the inverted S/N cube. 
	}
\end{figure*}
\begin{figure*}
	\begin{center}
	\includegraphics[trim=10 -30 20 40, width=170mm]{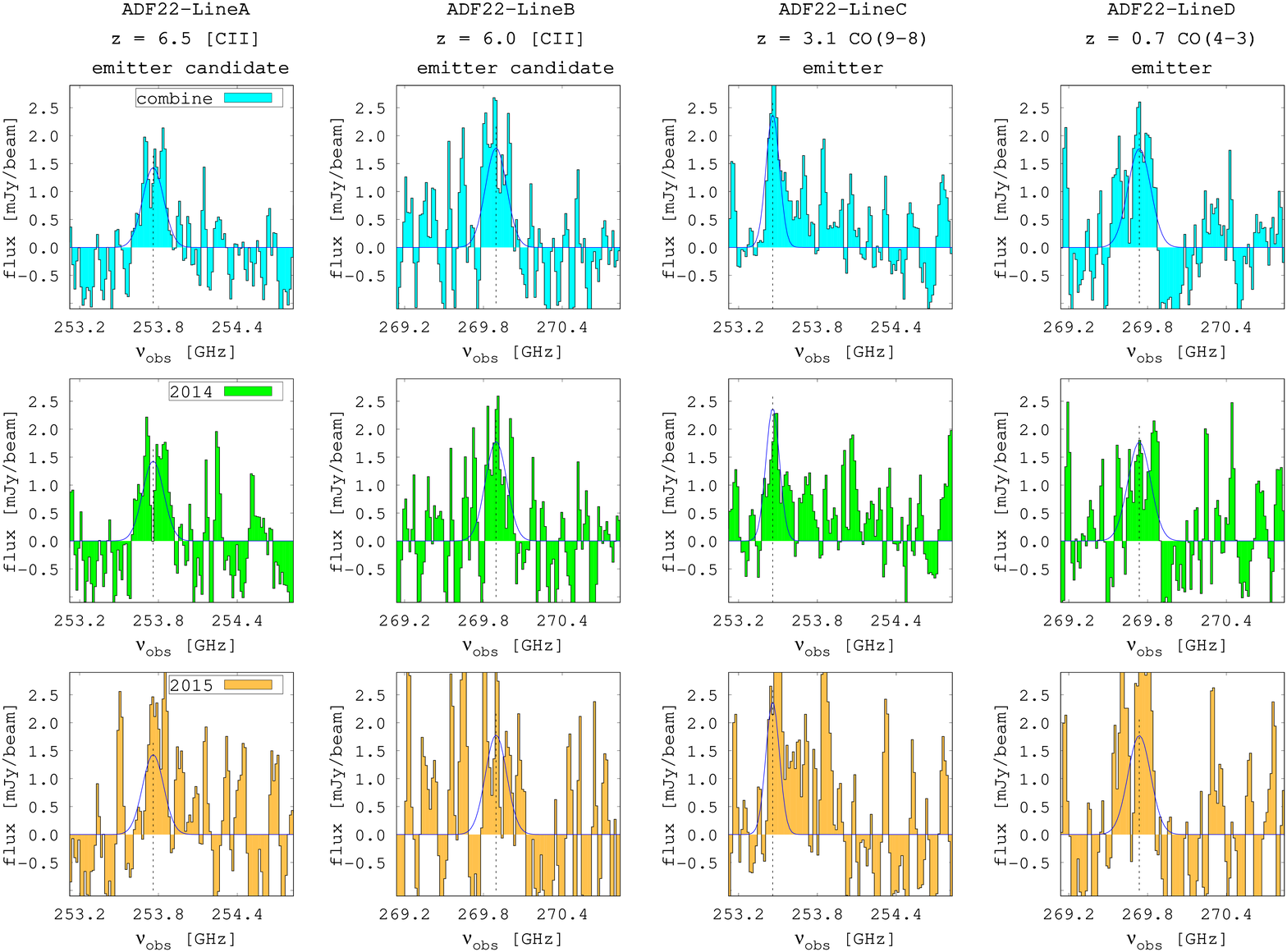}
	\end{center}
	\caption{
	The spectra of our four line emitter candidates .  
  	We use continuum-subtracted image data in original spectral resolution.  
    	Top panels show the combined data observed in 2014 and 2015, whereas
  	the middle and the bottom panels show separately the data observed in 2014 (green) 
  	and the data observed in 2015 (orange).  
  	The blue curve in each panel shows the best fit Gaussian profile for the combined data.
	ADF22-LineA and LineB are the candidate [\cii] emitters from our survey. 
	}
\end{figure*}
	We detect four line emitter candidates.  
	Hereafter, we call the two sources without optical, NIR and FIR counterparts ADF22-LineA and 
	ADF22-LineB.  Those with counterparts are dubbed as ADF-LineC and ADF-LineD.  
	The peak S/N ratio are $6.5 \sigma$, $6.2 \sigma$, $7.7 \sigma$ 
	and $6.5 \sigma$, for ADF22-LineA, B, C and D, respectively.  
	The first moment images of the candidates are shown in Figure A, and 
	their properties are shown in Table 2.  

	Figure 3 shows the cumulative number of positive and negative clumps as
        a function of S/N ratio.  
	Although the S/N ratios of ADF22-LineA, B and D are below $6 \sigma$ at the original
        spectral sampling, the lines are detected at $\geq 6 \sigma$ in the smoothed S/N cubes. 
	We compare the spectral line features of the emitter candidates 
	in different observation epochs, 2014 and 2015 (Figure 4).  
	Overall, the contiguous positive signals over a velocity range of $\geq$ 180 km s$^{-1}$ and
	the line features commonly seen suggest that the candidates are likely real sources.
	
	We note here that we also detect one clump with $6.3 \sigma$ in the inverted S/N cube, 
	and thus we would naively be concerned that one candidate with $6.2 \sigma$ could be
        a spurious source.
        However, the most-negative clump is actually detected in SPW 3,  
        where none of our four candidates is located. 
        We also find that datacubes with higher RMS value have higher maximum negative detection (Table 1).
	Since the datacubes in different SPWs have different properties, 
	the existence of the high-$\sigma$ negative clump in SPW 3 does not immediately
        impacts the confidence of our line emitter candidates.
	ADF22-LineB has a lower S/N ratio than ADF22-LineA, whereas it has non-negative $z'$ band 
	counterparts with $< 3 \sigma$ (see also Figure 5).  
	Velocity-gradient is also seen around ADF22-LineB (see also Figure A).  

\begin{figure*}
	\begin{center}
	\includegraphics[width=150mm, trim= 0 -10 40 10]{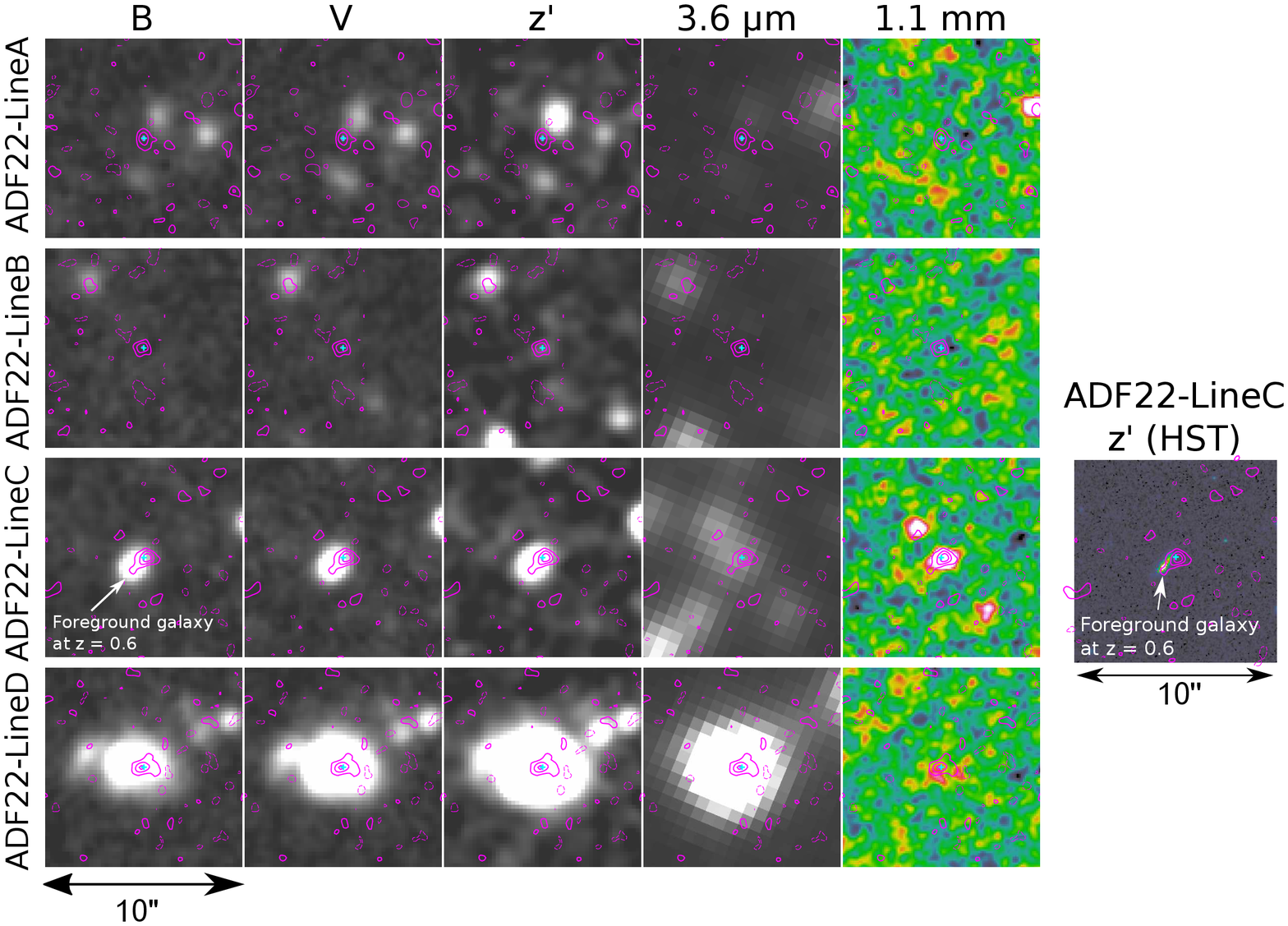}
	\end{center}
  	\caption{
	  Images of the four line-emitter candidates at different wavelength range
          from optical to 1.1 mm. 
	  In each panel, the cyan cross indicates the position of the candidate.  
	  The magenta contours show the 0th moment images of the line emission 
          with $2,~4,$ and $6 \sigma$, and the dashed contours with $-4$ and $-2~\sigma$.
	  We adopt a linear colour-scale for this figure.  
	  Counterparts are not found at the position of LineA and LineB at any wavelength. 
	  We also show a z' band image of LineC taken by {\it HST} Advanced Camera for Surveys 
	  I (F814W)-band in archive (PID 9760) in the rightmost panel. 
	}
\end{figure*}

\subsection[]{Line Identification}
	Figure 5 shows the images of the four candidates in $B$, $V$, $z'$, 3.6 $\rm \mu$m and 1.1 mm 
	wavebands.  We plot SED and model fit for ADF22-LineD in Figure B, and
	the measured photometry in the detected bands are given in Table 3.   
	The photometric redshift is estimated by using {\sc hyperz} \citep{bolzonella2000}. 
	We fit the SED templates by \citet{bruzual1993} to the spectral coverage from UV to 8 $\rm \mu m$, 
 	assuming a Calzetti dust extinction law \citep{calzetti2000}. 
	We also use SED templates from {\sc swire} library \citep{polletta2007}.  

	{\bf ADF22-LineA and B:}  
	We do not find any secure counterpart nor close sources within 2$''$ of the sources. 
	Therefore we regard LineA and LineB as good [\cii] emitter candidates.  
	
	{\bf ADF22-LineC:}  LineC very likely arises from the galaxy ADF22.4 reported in \citet{umehata2017}, 
	whose redshift is determined to be $z = 3.091$ 
	from far-infrared spectroscopic follow up observations (Umehata et al.\, in prep.).  
	Thus we identify ADF22-LineC as CO(9-8) line emission at $z = 3.091$.  
	In addition, the optical component near ADF22-LineC is a known galaxy at $z = 0.55$
	 \citep{kubo2015}, but we exclude possibility of  ADF22-LineC to be at $z = 0.55$ 
	 because there is no obvious line species observed at 1.1 mm.   
	LineC is also detected in X-ray \citep{lehmer2009}, 
	which may indicate that ADF22-LineC is an AGN-host galaxy. 
	Further details of the galaxy will be discussed in Umehata et al.\,(in prep.).  
\begin{figure*}
	\begin{center}
	\includegraphics[width=140mm, trim= 50 -20 120 30]{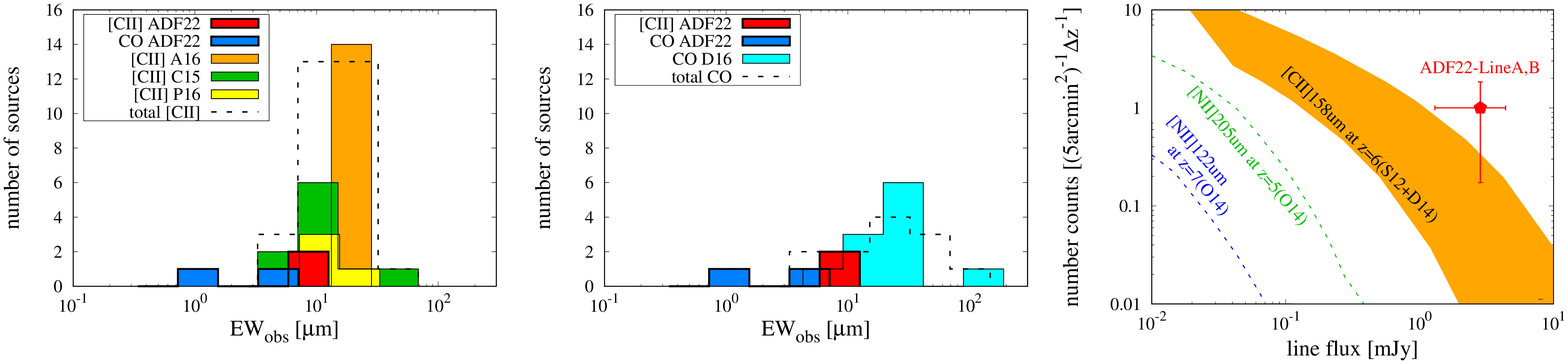}
	\end{center}
	\caption{ 
	  $Left~and~middle$: We compare the distribution of EWs in 0.9--1.3 mm observed frame: 
	    The ADF22 sources with high redshift [\cii ] emitters/candidates (left) 
	    and low-redshift CO emitters/candidates (middle) \citep{capak2015, aravena2016b, pentericci2016, decarli2016b}.  
	    Continuum upper limit is assumed to be $3 \sigma$.  The histograms except for the case of 
	    ADF22 CO emitters contain the data with lower limits.  
	    EW values of ADF22 [\cii] emitter candidates are comparable to other high-redshift [\cii ] emitters/candidates 
	    and  ADF22 CO emitters have in relatively lower values.  
	    For CO emitters, their transition are $j$ = 9--8 to 3--2, which are not corrected to fixed transition. 	    
	    ADF22 [\cii ] emitter candidates are consistent with the CO distribution using observational data available.  
	    $Right$: The cumulative number counts of [N{\sc ii}]-, and [\cii]-line emitters, 
	    where ${\rm \Delta}z$ is set to be the survey redshift range.  
	    Red point represents number density of this survey for one [C{\sc ii}] emitter candidate 
	    with $1\sigma$ error bar from the source number statistics \citep{gehrels1986}.  
	    Orange shaded region shows the [\cii] emitter number count estimated (converted) 
	    from the star formation rate function at $z = 6$ \citep{smit2012} and 
	    SFR-$L_{\rm [CII]}$ relation \citep{delooze2014}.  The velocity width is assumed to be 200 km s$^{-1}$. 
	    This simple model shows agreement to the observed number count.  
	    The dotted lines show [N{\sc ii}]-emitter number counts from Orsi et al. (2014), 
	    which are well below the observational result. 
	    We also note the [O{\sc iii}] number counts at $z \sim 12$ are below 
	    of the [NII]122$\rm \mu$m counts \citep{orsi2014}. 
	}
\end{figure*}

	{\bf ADF22-LineD:}  LineD is spatially consistent with the position of a tentatively detected 
	continuum source ADF22.21 reported in \citet{umehata2017}.  
	The result of SED fitting shows that reduced $\chi^2$ values becomes 
	minimum at $z \sim 0.6-0.8$ (Figure B left).  
	Interestingly, the SED is well fitted by that of Arp220 placed at $z \sim 0.7$ (Figure B right). 
	By searching for possible lines in this redshift range, 
	we conclude that ADF22-LineD is likely a CO(4-3) emitter at $z = 0.71$. 

\subsection[]{Possibility of other line emissions}
	Besides the [\cii] line emission, 
	there are also possibilities that ADF22-LineA and LineB are other emission
        line sources, such as $^{12}$CO line emission at $z \leq 3.1$, H$_2$O at $z \sim 1.9$ or $2.8$, 
        	[N{\sc ii}]205$\rm \mu m$ at $z \sim 4.6$, 
	[O{\sc i}]145$\rm \mu m$ at $z \sim 6.9$, 
	[N{\sc ii}]122$\rm \mu m$ at $z \sim 8.5$ 
	or [O{\sc iii}]88$\rm \mu m$ at $z \sim 12$ 
	\citep{swinbank2012, tamura2014, ono2014, decarli2016, aravena2016b}.  

	If ADF22-LineA and LineB are $^{12}$CO emitters, the number density
	is consistent with the result of ASPECS survey \citep{decarli2016} 
	and with semi-analytical/empirical predictions referred to the article 
	\citep{lagos2012, popping2016, vallini2016}.  
	Thus we cannot exclude the possibility of $^{12}$CO emitters by the discussion of detectability.  

	We compare the equivalent widths (EWs) in observed frame of the four candidates. 
	The estimated EWs are $>$ 8.6, $>$ 14.6, 1.1 and 7.3 $\rm \mu m$ for ADF22-LineA, B, C and D, 
	respectively, assuming $3 \sigma$ continuum flux limit.  
	ADF22-LineA and B have higher EW than the blindly detected $^{12}$CO emitters in our survey.  
	The left and middle panels of Figure 6 also show the distribution of the EWs in 0.9--1.3 mm
         observed frame of the four candidates, high-redshift [\cii] emitting LBGs and LAEs 
         \citep{capak2015, pentericci2016}, 
	[\cii] emitter candidates detected in ASPECS \citep{aravena2016b}, 
	$^{12}$CO emitter candidates detected in band 6 in ASPECS \citep{decarli2016b}. 
	The EWs of ADF22-LineA, B and other high-redshift [\cii] emitter/candidates 
	are comparable.  Given these information, we argue that ADF22-LineA and B are more likely 
	[\cii] emitters at $z = 6.5$ and $6.0$, rather than CO emitters at $z \leq 3.1$.  
	EW values of ADF22-LineA and B are comparable to those of the blindly detected CO emitters. 
	Further consideration by using forthcoming follow up observation and theoretical study will be needed to 
	yield any insight about the trends of EW distributions.  
    	As with ADF22-LineA and B, blindly detected line emitter candidates are expected to have often 
	no counterpart \citep{aravena2016b}. Thus it is important to study the EWs of a large sample of CO/[\cii] emitters.  
	We note that H$_2$O molecular lines are expected to have similar line flux to CO line
        emission in the submm band (e.g,\,\cite{rangwala2011,omont2013}), 
        and thus can be distinguished from high-redshift [\cii] emitters by comparing their EWs.  
\begin{figure*}
	\begin{center}
	\includegraphics[width=180mm, trim= 80 -10 0 0]{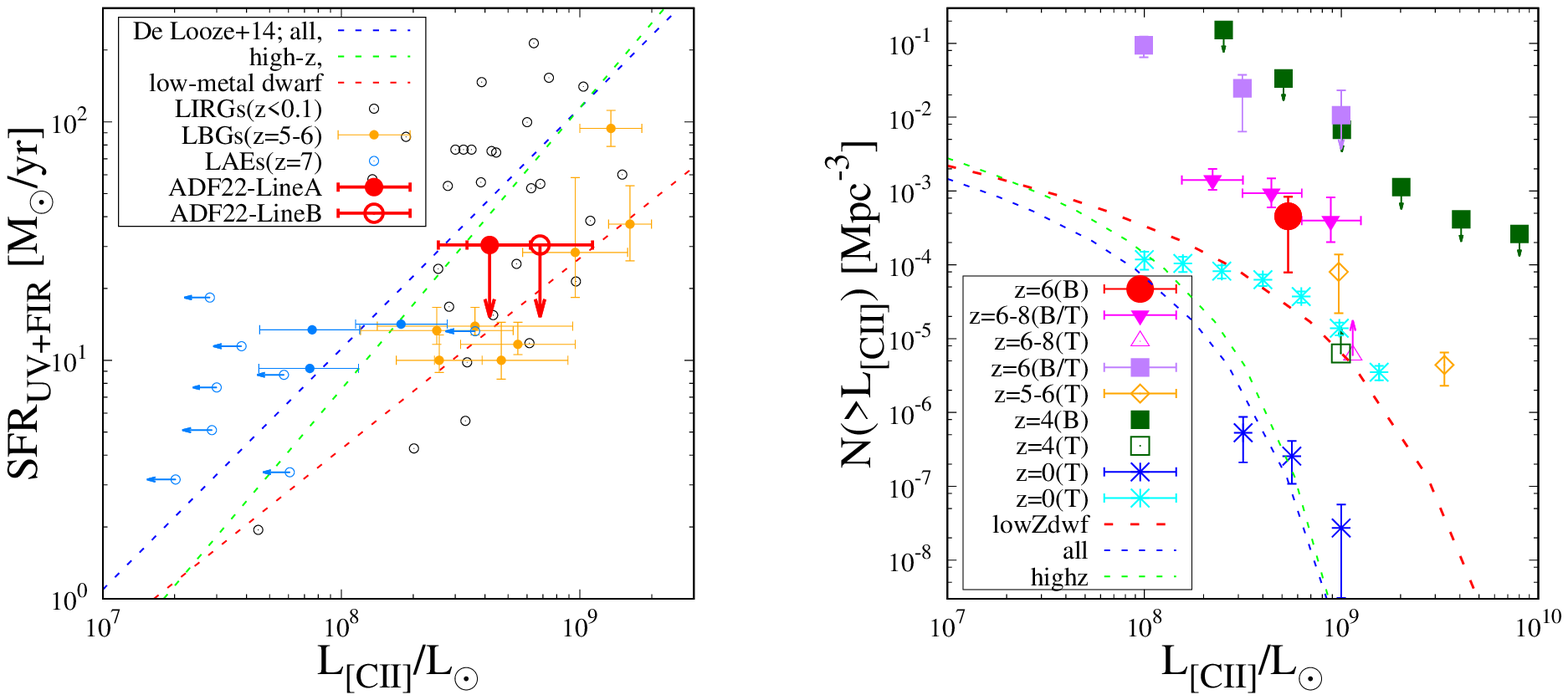}
	\end{center}
	\caption{
	$Left$: We compare the [\cii] emitter candidates with 
	other local luminous infrared galaxies (LIRGs) \citep{maiolino2009}, LBGs at $z=5.2-5.7$ \citep{capak2015}, 
	and LAEs at $z = 6.6-7.2$ \citep{pentericci2016} in the SFR - $L_{\rm [CII]}$ plane.  
	The dashed lines represent empirical SFR-$L_{\rm [CII]}$ relations by \citet{delooze2014}
        for different populations. 
        The SFR-$L_{\rm [CII]}$ relation for metal poor dwarfs is consistent with 
        the upper limit of SFR$_{\rm UV + FIR}$ of [\cii] emitter candidates.  
        $Right$: We summarize the redshift evolution of the [\cii] luminosity function ([\cii] LF)
        	\citep{swinbank2012, hemmati2017, matsuda2015, capak2015, miller2016, aravena2016b}. 
        The estimations from targeted observations are denoted by (T), 
        and blind surveys are denoted by (B).  
	We estimate [\cii] LF at $z = 6.2$ from blind detection on the assumption that 
	one of the two unconfirmed lines is [\cii] line at $z \sim 6$.  
	The error-bar on our point uses Gehrels (1986).  
	We also plot model of [\cii] LFs at $z = 6$ calibrated 
	by using the star formation rate function at $z = 6$ \citep{smit2012}
	and the SFR-$L_{\rm [CII]}$ relation by \citet{delooze2014}.
	The observational results at $z > 4$ show good agreement with the predicted LF for metal poor dwarfs.
	}
\end{figure*}

	[\cii ] luminosity, $L_{\rm [CII]}$ of ADF22-LineA and B is calculated by using luminosity distance $D_{\rm L}$, 
	observed frequency $\nu_{\rm o}$, velocity-integrated flux $S^{\rm v}$ (e.g.,\,\cite{carilli2013}), 
	\begin{equation}
	\frac{L_{\rm [CII]}}{L_\odot} = 1.04 \times 10^{-3} \left ( \frac{D_{\rm L}}{\rm Mpc} \right )^2 
	\frac{\nu_{\rm o}}{\rm GHz} \frac{S^{\rm v}}{\rm Jy~km~s^{-1}}. 
	\end{equation}	
	The estimated $L_{\rm [CII]}$ of $4-7 \times ~10^8~L_\odot$ is consistent with 
	the values of normal star-forming galaxies in the local universe 
	(e.g.,\,\cite{swinbank2012}), thus we do not consider the effect of [\cii] line deficit (e.g.,\,\cite{gracia2011,diaz2013}).  
	We then derive [\cii] luminosity function by using SFR-[\cii] luminosity relation \citep{delooze2014} 
	and SFR function at $z = 6$  \citep{smit2012}.   
	The right panel of Figure 6 shows that the detection of one [\cii ] emitter candidate in the survey area 
	is roughly consistent with the expected [\cii] number counts, 
	if we use the SFR-$L_{\rm [CII]}$ relation by \citet{delooze2014} 
	that is calibrated from observations of nearby low-metallicity dwarf galaxies (see also \S 5): 
	\begin{equation}
	\frac{\rm SFR_{\rm [CII]}}{\rm M_\odot yr^{-1}} =  10^{-5.73 \pm 0.32} 
		\left ( \frac{L_{\rm [CII]}}{\rm L_\odot}\right )^{0.80 \pm 0.05}.
	\end{equation} 
	
	We also plot the predicted number counts of [N{\sc ii}] 122$\rm \mu$m 
	and [N{\sc ii}] 205$\rm \mu m$ from the model of Orsi et al. (2014).  
	The predicted number count of [O{\sc iii}] 88$\rm \mu$m emission at $z \sim 12$ are
	lower than the [NII] 122$\rm \mu$m emission \citep{orsi2014}. 
	It is expected that such line emitters will not be found in our survey area.  
	From the discussion above, we assume ADF22-LineA and B to be [\cii ] emitter candidates. 

\section[]{Discussion}
\begin{figure*}
	\begin{center}
	\includegraphics[width=100mm, trim= 40 -20 50 0]{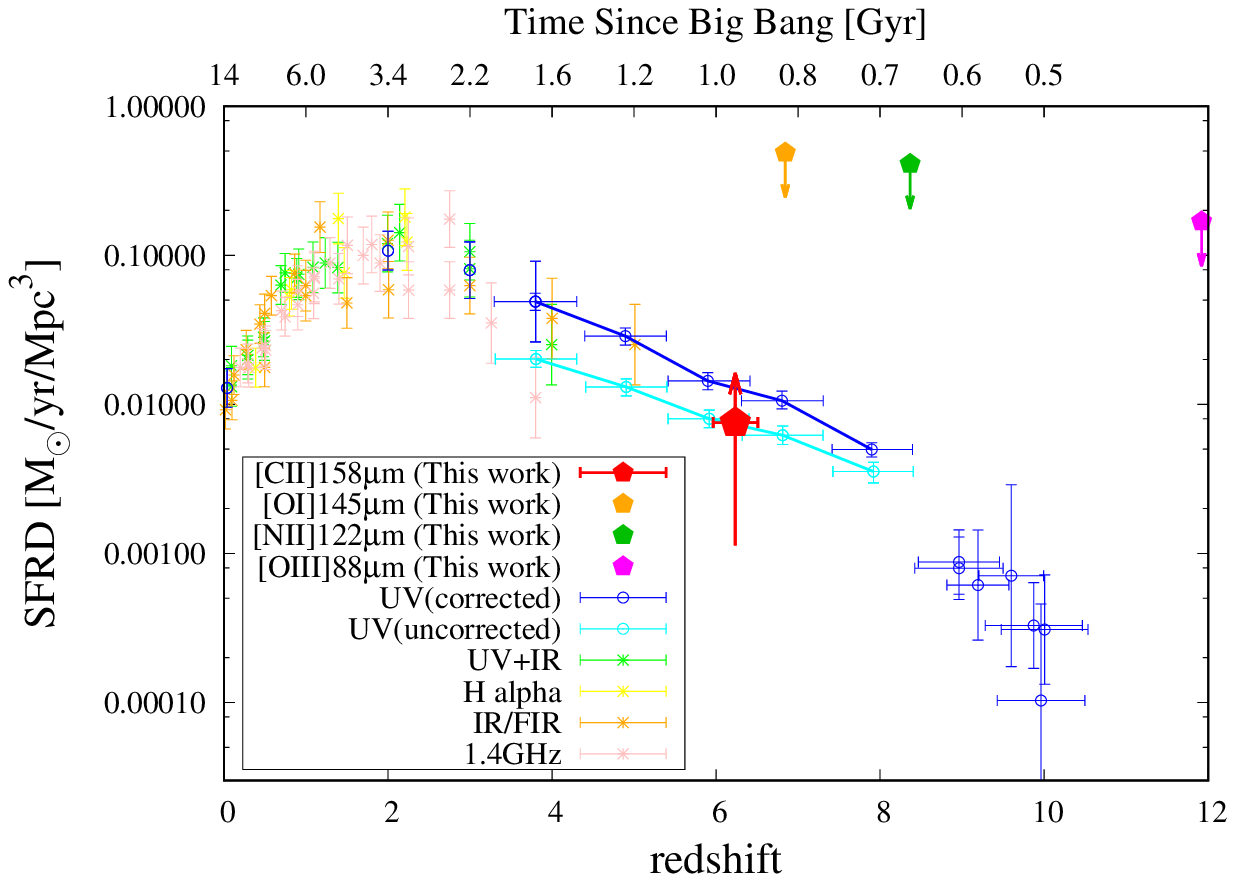}
	\end{center}
	\caption{
         The cosmic star-formation rate density (SFRD) 
        as a function of redshift.  
	We show lower limit of a [\cii] SFRD at $z = 6$ derived 
	from the mean SFR of the two [\cii] candidates divided by the survey volume.  
	We also estimate the upper limits for the un-detected fine-structure lines at $z = 7--12$.   
	We plot the SFRD estimated by rest-frame UV observations \citep{oesch2014, capak2015} 
	and other wave bands\citep {behroozi2013}.  
	The estimation of [\cii] SFRD from a conservatively selected one source
        is already consistent with the dust-uncorrected UV SFRD at $z = 6$. 
	}
\end{figure*}
	In order to discuss the cosmic star formation history, 
        we derive the SFRs of ADF22-LineA and ADF22-LineB assuming that they 
        are [\cii] emitters at $z = 6$.  
        We calculate total SFR by summing up the dust-uncorrected SFR$_{\rm UV}$ 
        and SFR$_{\rm IR}$,  SFR$_{\rm UV + IR}$(e.g.,\,\cite{buat2010}), 
        by using the following equations \citep{kennicutt1998}; 
	     \begin{eqnarray}
	       {\rm \frac{SFR_{UV}}{M_\odot ~yr^{-1}}} 
			&=& 7.8 \times 10^{-29} \frac{L_{\rm \nu}}{{\rm ergs~s^{-1}~Hz^{-1}}},\\
 	{\rm \frac{SFR_{IR}}{M_\odot ~yr^{-1}}} 
			&=& 2.5 \times 10^{-44} \frac{L_{\rm IR}}{{\rm ergs~s^{-1}}},
	\end{eqnarray}
	where L$_{\rm \nu}$ refers to the UV luminosity density in the wavelength range 1500-2800 \AA, 
	and L$_{\rm IR}$ refers to the IR luminosity integrated over 8-1000 $\rm \mu$m.  
	We estimate $L_{\rm IR}$ of ADF22-LineA, B and D 
	from the observed 1.1 mm continuum fluxes
	by using the SED fitting method of \citet{charyelbaz2001}.  
	Continuum upper limits of LineA and B are assumed to be $3 \sigma$.	
	$L_{\rm IR}$ of ADF22-LineC is referred to estimation of \citet{umehata2017}. 
	Upper limit of UV luminosity is estimated in \citet{nakamura2011}.  
        The obtained SFR$_{\rm UV + FIR}$ is $<$ 30 $\rm M_\odot yr^{-1}$, 
        being consistent with the SFR-$L_{\rm [CII]}$ relation of \citet{delooze2014} that is 
        calibrated by local low-metallicity dwarf galaxies (Figure 6 left).          
	We note that the relation calibrated by high-$z$ galaxies is considered to be applicable to 
	bright [\cii ] emitters with $>~10^9 \rm L_\odot$ \citep{delooze2014} 
	and exceeds the SFR$_{\rm UV + FIR}$ upper limit for the candidates in this survey.  
        The estimated SFR$_{\rm [CII]}$ from the low-metal dwarf relation are 
        13$^{+4}_{-5}$ and 20$^{+9}_{-10}$ M$_\odot$/yr for ADF22-LineA and LineB, respectively (see also Table 3).  
	
	We estimate the [\cii] luminosity function (LF) at $z \sim 6.2$ from only one source, because 
        one of the two [\cii] candidates has a relatively low-$\sigma$ and thus could possibly
        be a spurious source (see \S 4.1). 
        We show this result in Figure 7 and compare it to [\cii] LFs from previous studies.  
	The estimated [\cii] LFs at $z = 0.0$ \citep{swinbank2012, hemmati2017} 
	are derived from the follow up observation of the {\it IRAS} sources \citep{brauher2008} 
	or samples from the Great Observatories All-sky LIRG Survey \citep{diaz2013}. 
	We indicate the upper limit at $z = 4.4$ derived by \citet{matsuda2015} using ALMA Cycle 0 archive data,
	and the lower limit at $z = 4.4$ based on two serendipitous detection in ALESS survey \citep{swinbank2012}.  
	The estimation at $z = 5-6$ is derived from follow up observation by \citet{capak2015}.  
	We also indicate the estimation of over-dense region at $z = 6$ by \citet{miller2016} . 
	The constraint for [\cii] LF at $z = 6-8$ is provided by result of ASPECS \citep{aravena2016b}, 
	which is based on an assumption that all [\cii ] candidates are real [\cii] emitters. 
	As discussed In \S 4.2, We also derive the simple model of [\cii ] luminosity function at $z$ = 6 by 
	using SFR-$L_{\rm [CII]}$ relation \citep{delooze2014} and SFR function at $z$ = 6 \citep{smit2012}. 
        Our [\cii] LF model is close to our own observational result and the other studies, 
        whereas the estimated LFs based on the empirical relations for high-redshift, and for all galaxies
        \citep{delooze2014}, do not match the observational result at $z > 4$.  
        We note that if completeness of the detection is lower than unity, the estimated [\cii ] LF represents 
        the lower limit. 

	We calculate a conservative limit of [\cii] SFRD from the mean of the SFR$_{\rm [CII]}$ 
	of the two sources divided by the survey volume (Figure 8).  
	The derived [\cii] SFRD is $> 7.5 \times 10^{-3} \rm M_\odot yr^{-1}  Mpc^{-3}$. 
	Interestingly, this is close to the dust-uncorrected UV SFRD at $z = 6.2$.  
	The contribution of the only one [\cii] emitter with 
	faint UV and dust emission to the cosmic SFRD might already constitute a major contribution.  
	The result may imply the existence of the untraceable 
	component of the SFRD by rest-UV. 
	In order to confirm the truth of this, the estimation of a faint end slope of the [\cii ] LF would be crucial.  
	
	In figure 8, we also derive upper limits of SFRDs at $z = 7-12$ from 
	the non-detections of [O{\sc i}] 145$\rm\mu$m,  [N{\sc ii}] 122$\rm \mu$m [OIII] 88$\rm \mu$m lines 
	in our search as discussed in \S 4.3. The SFRs are calculated from line luminosities 
	by using observational relations estimated by \citet{farrah2013}.  
	This result demonstrates that line survey enables us to estimate SFRDs at multiple redshifts at once. 
	
	There are a few possible mechanisms for the [\cii ] line emission to be particularly intense 
	relative to FIR and UV emission. For example, it can be caused by high far-UV radiation from massive, 
	young stars in the early universe (e.g.,\,\cite{wolfire1995}).  
	The environment of a low metallicity and a low dust-to-gas ratio can also cause 
	enhancement of [\cii ] radiative cooling \citep{wolfire1995,capak2015}.
        In particular, the low dust-to-metal environment 
	may not only enhance [\cii ] line emission but also weaken dust continuum emission
        \citep{inoue2003, asano2014}.  
	Hot dust dominates the short-wavelength portion of the SED \citep{casey2014, zhou2016}, 
	making the dust continuum at long-wavelengths to be relatively suppressed.  
	The size distribution of dust grains also affect faint FIR continuum \citep{takeuchi2003, takeuchi2005b}.  
        Altogether, observations in the submm-band can provide invaluable information on the physical
        properties of high-redshift galaxies.  Future deep submm surveys will enable us to understand the formation of
        galaxies and to probe the early cosmic star-formation history. 

\section*{Summary}	
	We search millimeter line emitters by using 1.1 mm ADF22 survey data taken in ALMA Cycle2.  
	Our newly constructed method for line search worked for detecting two CO emitters at $z = 0.7$ and $3.1$ 
	and two [\cii ] emitter candidates at $z = 6.0$ and $6.5$ with $> 6 \sigma$. 
	[\cii ] emitter candidates are faint in all counterparts. 
	The line species of the CO emitters are identified by SED fitting or spectral follow up observation.  
	For [\cii] emitter candidates, the possibility of other line emissions are excluded by discussion 
	about number counts, line ratio and EWs.  
	Since one spurious source is possibly contaminated with the candidates, 
	we assume at least one of the two candidates to be a real [\cii] emitter. 
	We constrain $z = 6$ [\cii] LF for one source and found that the [\cii ] LFs at $z > 4$ 
	show good agreement with the predicted LF by using 
	SFR-$L_{\rm [CII]}$ relation calibrated by local metal poor dwarfs. 
	We also found that estimated [\cii ]-based SFR are consistent with upper limit of total SFR 
	if we use the SFR-$L_{\rm [CII]}$ relation for local metal poor dwarfs. 
	We estimate a conservative limit of [\cii] SFRD at $z = 6.2$ for one source, 
	which is close to the dust-uncorrected UV SFRD at $z = 6.2$. 
	The results might be imply that mm/submm line survey is a powerful probe 
	to estimate untraceable SFRD component from rest-UV observation at high-redshift. 
	The constrain for faint end slope of [\cii ] LF from further line survey and FIR/UV follow-up observation 
	will give us the truth of such implication and detailed picture of cosmic star-formation history. 

\begin{landscape}
\begin{table}[th]
	\caption{\sc Properties of datacube of Four Spectral Windows.}
	\centering
	\begin{threeparttable}
	\begin{tabular}{c c c c c c c c c c}\hline
	SPW & $\nu_{\rm obs}\tnote{(1)}$ & $z_{\rm [CII]}$\tnote{(2)} & d$v$\tnote{(3)}
	& angular resolution\tnote{(4)} & RMS of original data\tnote{(5)}  
		& \# of net & \# of  matched& max positive & max negative\\
	ID & [GHz] & & [km s$^{-1}$] & [arcmin] &[mJy/beam] & clumps\tnote{(6)} & clumps & S/N ratio & S/N ratio \\ \hline \hline 
	0	& 253.12 -- 254.83 & 6.458 -- 6.508 & 18.3  & 0.67, 0.53, 1.09 & 0.7, 0.9, 1.2, 
		& 25 /18 & 9 /10 & 7.77$\sigma$ (10\tnote{(7)}) & 5.70$\sigma$ (2) \\ \hline
	1	& 255.14 -- 256.83 & 6.400 -- 6.449 & 18.1  & 0.68, 0.54, 1.11 & 0.7, 0.8, 1.1, 
  		& 6 /4 & 5 /3 & 5.73$\sigma$ (6) & 5.81$\sigma$ (0) \\ \hline	
	2	& 269.14 -- 270.84 & 6.017 -- 6.062 & 17.2  & 0.62, 0.49, 1.02 & 0.8, 1.0, 1.4, 
	 	& 18/14 & 7 /6 & 6.51$\sigma$ (21) & 6.05$\sigma$ (6) \\ \hline
	3	& 271.14 -- 272.84 & 5.966 -- 6.009 &17.1 & 0.62, 0.49, 1.01 & 0.9, 1.2, 1.6, 
		& 10/21 & 5 /10 & 5.99$\sigma$ (21) & 6.30$\sigma$ (8)  \\ \hline 
       	\end{tabular}
	\begin{tablenotes}
	\item[(1)] Observed frequency range we use for the search.  
	\item[(2)]  [\cii] redshift range corresponding to the observed frequency range.  
	\item[(3)]  Mean velocity corresponding to an interval of slices.  
	\item[(4)] Mean angular resolution. Column shows combined data, 2014 data and 2015 data, respectively.  
	\item[(5)]  1 $\sigma$ sensitivity at 36 km s$^{-1}$ spectral resolution calculated 
	by using primary beam corrected data.  Column shows in the same manner to (4).  
	\item[(6)]  Number of clumps detected by {\sc clumpfind} \citep{williams1994} 
	in original/inverted S/N cubes.  
	\item[(7)]  Numbers in bracket represent the size of slices smoothed in spectral
        domain of the S/N cube.  
	\end{tablenotes}
	\end{threeparttable}
\end{table}

\begin{table}[th]
 	\caption{\sc Measured and  derived candidate properties.}
	\centering
	\begin{threeparttable}
  	\begin{tabular}{c c c c c c c c c c c c c}\hline
	ADF22 ID & Name & $\nu_{\rm peak}$ &  S$_{\rm peak}$/N\tnote{(1)} & Smoothing\tnote{(2)}
		& $S_{\rm 1.1mm}$\tnote{(3)} & $S_{\rm line}$\tnote{(4)} & FWHM\tnote{(5)} & EW$_{\rm obs}$ 
		& $z$\tnote{(6)} & log$_{10}L_{\rm line}$\tnote{(7)} & log$_{10}L_{\rm FIR}$\tnote{(8)} \\
	& (J2000)  & [GHz] & [$\sigma$] & [km s$^{-1}$] & [mJy] & [Jy km s$^{-1}$] 
	& [km s$^{-1}$] & [$\rm \mu m$] & & [$L_{\odot}$] & [$L_{\odot}$]\\ \hline \hline
	ADF22-LineA & ALMAJ221737.43+001710.7 & 253.79 & 6.5 & 220 (12) 
		& $<$ 0.2 & 0.4$\pm 0.1$ & 220$\pm 40$ & $>$ 8.6 &
		6.489 & 8.6$_{-0.2}^{+0.2}$ & $<$ 11.4 \\ \hline 
	ADF22-LineB & ALMAJ221731.95+001820.3 & 269.92 & 6.2 & 258 (15)
		& $<$ 0.2 & 0.7$\pm 0.1$ & 220$\pm 45$ & $>$ 14.6 
		& 6.041 & 8.8 $_{-0.3}^{+0.2}$ &$<$ 11.4 \\ \hline 
	ADF22-LineC$$ & ALMAJ221736.97+001820.8 & 253.49 & 7.8 & 220 (12)
		& 2.0$\pm 0.1$ & 0.5$\pm 0.1$ &140$\pm 20$ & 1.1 $^{0.2}_{-0.2}$
		& 3.091 & 8.0 $_{-0.2}^{+0.2}$ & 12.6 $^{+0.2}_{-0.1}$ \\ \hline
	ADF22-LineD & ALMAJ221733.07+001718.8 & 269.70 & 6.5 & 361 (21)
		& 0.7$\pm 0.2$  & 1.0$\pm 0.1$ & 240$\pm 40$ & 7.3$^{+ 5}_{- 2}$ 
		& 0.709 & 6.7 $_{-0.3}^{+0.2}$ & 11.9 $^{+0.2}_{-0.2}$\\ \hline 
	\end{tabular}
	\begin{tablenotes}
	\item{(1)} The maximum S/N ratio of the clump in all S/N cubes.  
	\item{(2)} The smoothing window of the datacube with maximum positive S/N ratio. 
	Numbers in bracket represent the corresponding size of slices. 
	\item{(3)} 3$\sigma$ upper limit is estimated in \citet{umehata2017}.  
	\item{(4)} The integrated line flux estimated by {\sc imfit}.  
	\item{(5)}  Full width half maximum derived by gaussian fit. 
	\item{(6)}  The redshift derived from $\nu_{\rm peak}$. The uncertainty is $O(10^{-3})$.  
	\item{(7)}  Line emission of  [\cii] 1900.543 GHz, [\cii], CO(9-8) 1036.912 GHz and 
		CO(4-3) 461.041 GHz, respectively. 
	\item{(8)}  We estimate the $L_{\rm FIR}$ of ADF22-LineA, B and D by using 
		the SED fitting method of \citet{charyelbaz2001}.  
		$L_{\rm FIR}$ of ADF22-LineC is referred to estimation of \citet{umehata2017}. 
	\end{tablenotes}
	\end{threeparttable}
\end{table}

\end{landscape}

\begin{table*}
  	\caption{\sc Photometry.}
	\centering
	\begin{threeparttable}
	\begin{tabular}{c c c c c c c c}\hline
	ADF22 ID\tnote{ } & $u^*$ & $B$ & $V$ & $R$ & $i'$ & $z'$ & $J$ \\  \hline \hline
	ADF22-LineA,B & $>$ 26.6 & $>$ 27.0 & $>$ 27.1 & $>$ 27.2 & $>$ 26.9 & $>$ 26.2 & $>$ 24.2 \\ \hline
	ADF22-LineD & 23.84$\pm 0.05$ &23.46$\pm 0.03$ & 22.86$\pm 0.01$ & 22.08$\pm 0.01$ 
		& 21.47$\pm 0.01$ & 21.13$\pm 0.01$  & 19.79$\pm 0.01$ \\ \hline
	\\ \\
	\end{tabular}
						
	\begin{tabular}{c c c c c c c c}\hline
	ADF22 ID & $H$ & $K_s$ & $3.6 \rm \mu m$& $4.5 \rm \mu m$ 
		& $5.8 \rm \mu m$ & $8.0 \rm \mu m$ & $24 \rm \mu m$\\ \hline \hline
	ADF22-LineA,B & $>$ 24.1 & $>$ 24.1 & $>$ 24.9 & $>$ 24.1 & $>$ 22.0 & $>$ 21.6 & $>$ 28.0 \\ \hline
	ADF22-LineD & 20.21$\pm 0.02$ & 19.39$\pm 0.01$ & 19.49$\pm 0.01$ & 19.88$\pm 0.02$ 
		& 19.81$\pm 0.05$ & 20.04$\pm 0.10$ & 16.46$\pm 0.10$\\ \hline
	\end{tabular}
	\begin{tablenotes}
	\item[]~~{\sc Note. ---} 
	All unit of photometry flux is AB magnitude.  Upper limits are given by 3$\sigma$.  
	The  CFHT MegaCam $u^*$ (in archive, P.I. Cowie, see also Matsuda et al. 2004).
	Subaru Suprime-Cam $B$, $V$, $R$, $i'$, $z'$ band \citep{matsuda2004}, 
	Subaru MOIRCS $J$, $H$, $K$s (Uchimoto et al. 2012) and the Spitzer IRAC 3.6, 4.5, 5.8, 8.0 $\rm\mu$m 
	(Webb et al. 2009)-band photometry of the line emitters. 
	The PSF difference in$u^*$ $\sim$ 8.0 $\rm \mu$m are corrected following Kubo et al. (2013). 
	The 24 $\rm \mu$m photometry flux is calculated over a 2$''$ diameter aperture. 
	\end{tablenotes}
	\end{threeparttable}
\end{table*}

\section*{Funding}
	This work was supported by the ALMA Japan Research Grant of National Astronomical 
	Observatory of Japan (NAOJ)  Chile Observatory, 
	NAOJ-ALMA-0071 and NAOJ-ALMA-0160. 
	NHH was supported by the grant of NAOJ Visiting Fellow Program supported 
	by the Research Coordination Committee, National Astronomical Observatory of Japan (NAOJ) and 
	by funding from Foundation for Promotion of Astronomy.
	IRS acknowledges support from STFC (ST/L00075X/1), 
	the ERC Advanced Grant DUSTYGAL (321334) 
	and a Royal Society Wolfson Research Merit Award.  
	
\section*{Acknowledgements}
	We are grateful to the referee R. Maiolino for his useful comments and suggestions.  
	The authors wish to thank A. Sternberg, E. Seaquist, L. Yao, M. Oguri, H. Nagai, I. Shimizu, K. Mawatari, T. Saito. 
	This paper makes use of the following ALMA data: ADS/JAO.ALMA\#2013.1.00162.S.  
	Data analysis ware carried out on common 
	use data analysis computer system at the Astronomy Data Centre, ADC, of the NAOJ.
	IRAC data was reduced and provided by J. Huang.  
	ALMA is a partnership of ESO (representing its member states), NSF (USA) and NINS (Japan), 
	together with NRC (Canada) and NSC and ASIAA (Taiwan) and KASI (Republic of Korea), 
	in cooperation with the Republic of Chile. The Joint ALMA Observatory is operated by ESO, 
	AUI/NRAO and NAOJ.  

\appendix
\renewcommand{\thefigure}{\Alph{figure}}
\setcounter{figure}{0}

\begin{figure*}
	\begin{center}
	\includegraphics[trim=10 0 20 0, width=150mm]{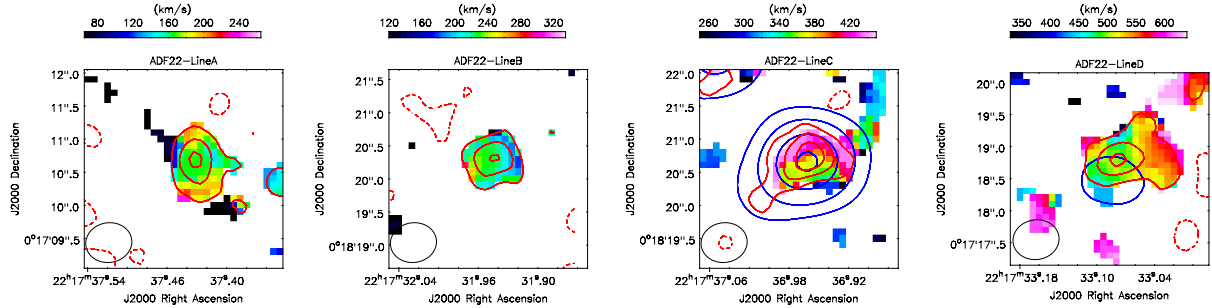}
	\end{center}
	\caption{
	We plot the continuum-subtracted first moment maps of the four candidates.  
  	The images are made by using the spectrally smoothed data.  
  	The colormaps show the projected velocity field with $>~2\sigma$ around each target.  
	The velocity is measured with respect to the center of each SPW.  
      	The red contours denote -2, 2, 4 and 6$\sigma$ with negative contours shown by dashed line
	whereas the blue contours show the S/N ratio of continuum, with 4, 10, 20, 30$\sigma$.  
  	ADF22-LineA and LineB are not detected in 1.1 mm continuum.  
  	The beam size is shown at the bottom-left corner of each panel.  
	}
\end{figure*}

\begin{figure*}
	\begin{center}
	\includegraphics[width=160mm, trim= 10 -20 40 40]{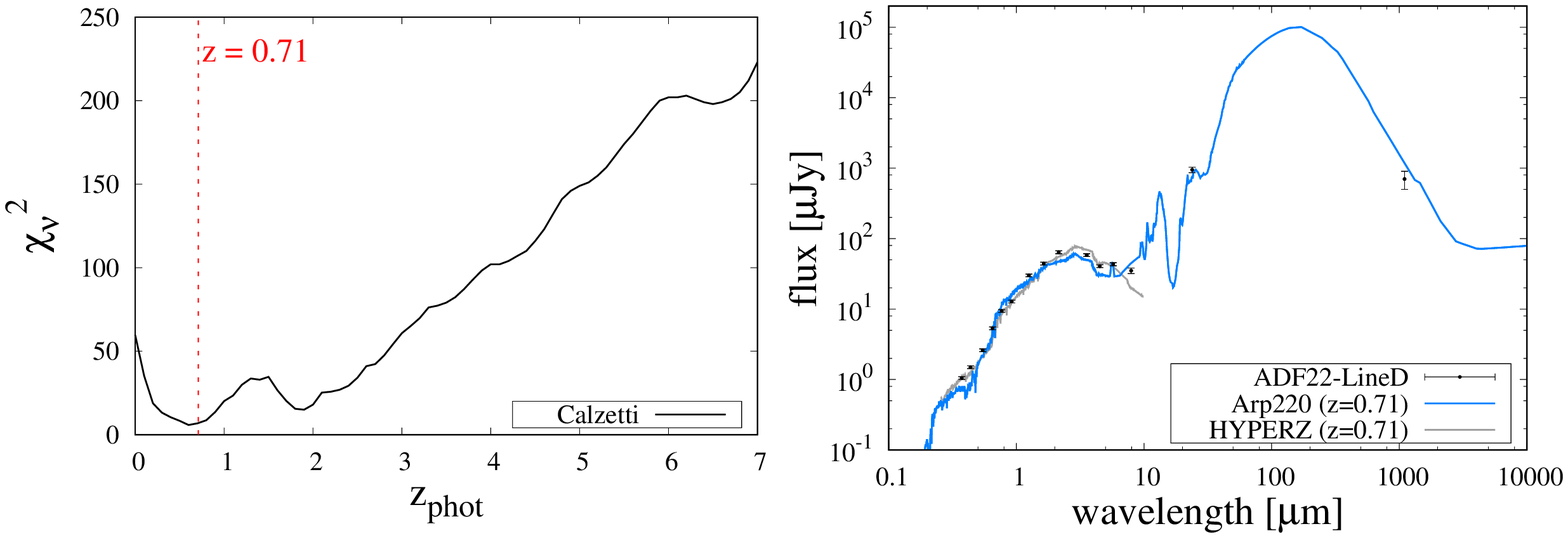}
	\end{center}
	\caption{
	$Left:$ The estimated reduced $\chi^2$ as a function of photometric redshift for the ADF22-LineD (right).  
	The $\chi^2_\nu$ value becomes minimum at $z=0.6-0.8$.  
	$Right:$ The spectral energy distribution (SED) of ADF22-LineD (right).  
	The SED is well fitted by that of Arp220 placed at $z \sim 0.71$.  
	 }
 \end{figure*}


\begin{thebibliography}{}
\bibitem[Asano et al.(2014)]{asano2014} 
Asano, R.~S., Takeuchi, T.~T., Hirashita, H., \& Nozawa, T.\ 2014, \mnras, 440, 134 

\bibitem[Aravena et al.(2016a)]{aravena2016a} 
Aravena, M., Decarli, R., Walter, F., et al.\ 2016, \apj, 833, 68

\bibitem[Aravena et al.(2016b)]{aravena2016b} 
Aravena, M., Decarli, R., Walter, F., et al.\ 2016, \apj, 833, 71 

\bibitem[Barger et al.(2012)]{barger2012} 
Barger, A.~J., Wang, W.-H., Cowie, L.~L., et al.\ 2012, \apj, 761, 89

\bibitem[Behroozi et al.(2013)]{behroozi2013} 
Behroozi, P.~S., Wechsler, R.~H., \& Conroy, C.\ 2013, \apj, 770, 57 

\bibitem[Blain et al.(1999)]{blain1999} 
Blain, A.~W., Smail, I., Ivison, R.~J., \& Kneib, J.-P.\ 1999, \mnras, 302, 632 

\bibitem[Blain et al.(2002)]{blain2002} 
Blain, A.~W., Smail, I., Ivison, R.~J., Kneib, J.-P., \& Frayer, D.~T.\ 2002, 
\physrep, 369, 111 

\bibitem[Bolzonella et al.(2000)]{bolzonella2000} 
Bolzonella, M., Miralles, J.-M., \& Pell{\'o}, R.\ 2000, \aap, 363, 476

\bibitem[Bouwens et al.(2012)]{bouwens2012} 
Bouwens, R.~J., Illingworth, G.~D., Oesch, P.~A., et al.\ 2012, \apj, 754, 83

\bibitem[Bouwens et al.(2016)]{bouwens2016} 
Bouwens, R.~J., Aravena, M., Decarli, R., et al.\ 2016, \apj, 833, 72 

\bibitem[Brauher et al.(2008)]{brauher2008} 
Brauher, J.~R., Dale, D.~A., \& Helou, G.\ 2008, \apjs, 178, 280-301 

\bibitem[Bruzual~\& Charlot(1993)]{bruzual1993}
Bruzual A., G., \& Charlot, S.\ 1993, \apj, 405, 538

\bibitem[Buat et al.(2010)]{buat2010} 
Buat, V., Giovannoli, E., Burgarella, D., et al.\ 2010, \mnras, 409, L1

\bibitem[Burgarella et al.(2013)]{burgarella2013} 
Burgarella D., et al., 2013, A\&A, 554, A70 

\bibitem[Calzetti et al.(2000)]{calzetti2000} 
Calzetti, D., Armus, L., Bohlin, R.~C., et al.\ 2000, \apj, 533, 682 

\bibitem[Carniani et al.(2015)]{carniani2015} 
Carniani, S., Maiolino, R., De Zotti, G., et al.\ 2015, \aap, 584, A78 

\bibitem[Carniani et al.(2017)]{carniani2017} 
Carniani, S., Maiolino, R., Pallottini, A., et al.\ 2017, arXiv:1701.03468

\bibitem[Casey et al.(2014)]{casey2014} 
Casey, C.~M., Narayanan, D., \& Cooray, A.\ 2014, \physrep, 541, 45 

\bibitem[Capak et al.(2015)]{capak2015} 
Capak, P.~L., Carilli, C., Jones, G., et al.\ 2015, Nature, 522, 455 

\bibitem[Carilli \& Walter(2013)]{carilli2013} 
Carilli, C.~L., \& Walter, F.\ 2013, \araa, 51, 105 

\bibitem[Chabrier(2003)]{chabrier2003} 
Chabrier, G.\ 2003, \pasp, 115, 763 

\bibitem[Chary \& Elbaz(2001)]{charyelbaz2001} 
Chary, R., \& Elbaz, D.\ 2001, \apj, 556, 562 

\bibitem[Chen et al.(2016)]{chen2016} 
Chen, C.-C., Smail, I., Swinbank, A.~M., et al.\ 2016, \apj, 831, 91 

\bibitem[Colbert et al.(1999)]{colbert1999} 
Colbert J.~W., et al., 1999, ApJ, 511, 721 

\bibitem[Cox et al.(2011)]{cox2011} 
Cox P., et al., 2011, ApJ, 740, 63 

\bibitem[da Cunha et al.(2013)]{cunha2013} 
da Cunha, E., Walter, F., Decarli, R., et al.\ 2013, \apj, 765, 9

\bibitem[Decarli et al.(2014)]{decarli2014} 
Decarli, R., Walter, F., Carilli, C., et al.\ 2014, \apj, 782, 78 

\bibitem[Decarli et al.(2016a)]{decarli2016} 
Decarli, R., Walter, F., Aravena, M., et al.\ 2016, \apj, 833, 69

\bibitem[Decarli et al.(2016b)]{decarli2016b} 
Decarli, R., Walter, F., Aravena, M., et al.\ 2016, \apj, 833, 70

\bibitem[De Looze et al.(2011)]{delooze2011} 
De Looze I., Baes M., Bendo G.~J., Cortese L., Fritz J., 2011, MNRAS, 416, 2712 

\bibitem[De Looze et al.(2014)]{delooze2014} 
De Looze, I., Cormier, D., Lebouteiller, V., et al.\ 2014, \aap, 568, A62

\bibitem[D{\'{\i}}az-Santos et al.(2013)]{diaz2013} 
D{\'{\i}}az-Santos, T., Armus, L., Charmandaris, V., et al.\ 2013, \apj, 774, 68

\bibitem[D{\'{\i}}az-Santos et al.(2016)]{diaz2016} 
D{\'{\i}}az-Santos, T., Assef, R.~J., Blain, A.~W., et al.\ 2016, \apjl, 816, L6 

\bibitem[Dunlop et al.(2017)]{dunlop2017} 
Dunlop, J.~S., McLure, R.~J., Biggs, A.~D., et al.\ 2017, \mnras, 466, 861

\bibitem[Farrah et al.(2013)]{farrah2013} 
Farrah, D., Lebouteiller, V., Spoon, H.~W.~W., et al.\ 2013, \apj, 776, 38

\bibitem[Fujimoto et al.(2016)]{fujimoto2016} 
Fujimoto, S., Ouchi, M., Ono, Y., et al.\ 2016, \apjs, 222, 1 

\bibitem[Geach \& Papadopoulos(2012)]{geach2012} 
Geach, J.~E., \& Papadopoulos, P.~P.\ 2012, \apj, 757, 156 

\bibitem[Gehrels(1986)]{gehrels1986} 
Gehrels N., 1986, ApJ, 303, 336 

\bibitem[Graci{\'a}-Carpio et al.(2011)]{gracia2011} 
Graci{\'a}-Carpio, J., Sturm, E., Hailey-Dunsheath, S., et al.\ 2011, \apjl, 728, L7

\bibitem[Gruppioni et al.(2013)]{gruppioni2013} 
Gruppioni, C., Pozzi, F., Rodighiero, G., et al.\ 2013, \mnras, 432, 23 

\bibitem[Hainline et al.(2009)]{hainline2009} 
Hainline, L.~J., Blain, A.~W., Smail, I., et al.\ 2009, \apj, 699, 1610

\bibitem[Hemmati et al.(2017)]{hemmati2017} 
Hemmati, S., Yan, L., Diaz-Santos, T., et al.\ 2017, \apj, 834, 36 

\bibitem[Hatsukade et al.(2016)]{hatsukade2016} 
Hatsukade, B., Kohno, K., Umehata, H., et al.\ 2016, \pasj, 68, 36 

\bibitem[Hayashino et al.(2004)]{hayashino2004} 
Hayashino, T., Matsuda, Y., Tamura, H., et al.\ 2004, \aj, 128, 2073

\bibitem[Herrera-Camus et al.(2016)]{herrera2016} 
Herrera-Camus, R., Bolatto, A., Smith, J.~D., et al.\ 2016, \apj, 826, 175

\bibitem[Hollenbach \& McKee(1989)]{hollenbach1989} 
Hollenbach, D., \& McKee, C.~F.\ 1989, \apj, 342, 306

\bibitem[Inoue(2003)]{inoue2003} 
Inoue, A.~K.\ 2003, \pasj, 55, 901

\bibitem[Inoue et al.(2014)]{inoue2014} 
Inoue, A.~K., Shimizu, I., Tamura, Y., et al.\ 2014, \apjl, 780, L18

\bibitem[Inoue et al.(2016)]{inoue2016} 
Inoue, A.~K., Tamura, Y., Matsuo, H., et al.\ 2016, Science, 352, 1559

\bibitem[Iono et al.(2006)]{iono2006} 
Iono, D., Yun, M.~S., Elvis, M., et al.\ 2006, \apjl, 645, L97

\bibitem[Kapala et al.(2015)]{kapala2015} 
Kapala, M.~J., Sandstrom, K., Groves, B., et al.\ 2015, \apj, 798, 24

\bibitem[Kaufman et al.(1999)]{kaufman1999} 
Kaufman, M.~J., Wolfire, M.~G., Hollenbach, D.~J., \& Luhman, M.~L.\ 1999, 
ApJ, 527, 795 

\bibitem[Kennicutt(1998)]{kennicutt1998} 
Kennicutt, R.~C., Jr.\ 1998, \apj, 498, 541

\bibitem[Kousai (2011)]{kousai2011} 
Kousai, K, \ 2011, Ph.D. Thesis, Tohoku University.

\bibitem[Kroupa(2001)]{kroupa2001} 
Kroupa, P.\ 2001, \mnras, 322, 231

\bibitem[Kubo et al.(2013)]{kubo2013} 
Kubo, M., Uchimoto, Y.~K., Yamada, T., et al.\ 2013, \apj, 778, 170 

\bibitem[Kubo et al.(2015)]{kubo2015} 
Kubo, M., Yamada, T., Ichikawa, T., et al.\ 2015, \apj, 799, 38 

\bibitem[Lagos et al.(2012)]{lagos2012} 
Lagos, C.~d.~P., Bayet, E., Baugh, C.~M., et al.\ 2012, \mnras, 426, 2142 

\bibitem[Lehmer et al.(2009)]{lehmer2009} 
Lehmer, B.~D., Alexander, D.~M., Geach, J.~E., et al.\ 2009, \apj, 691, 687 

\bibitem[Madau \& Dickinson(2014)]{madau2014} 
Madau, P., \& Dickinson, M.\ 2014, ARA\&A, 52, 415

\bibitem[Maiolino et al.(2005)]{maiolino2005} 
Maiolino R., et al., 2005, A\&A, 440, L51 

\bibitem[Maiolino et al.(2009)]{maiolino2009} 
Maiolino R., Caselli P., Nagao T., Walmsley M., De Breuck C., Meneghetti M., 2009, 
A\&A, 500, L1 

\bibitem[Maiolino et al.(2012)]{maiolino2012} 
Maiolino R., et al., 2012, MNRAS, 425, L66 

\bibitem[Maiolino et al.(2005)]{maiolino2005} 
Maiolino, R., Cox, P., Caselli, P., et al.\ 2005, \aap, 440, L51 

\bibitem[Maiolino et al.(2015)]{maiolino2015} 
Maiolino, R., Carniani, S., Fontana, A., et al.\ 2015, \mnras, 452, 54

\bibitem[Matsuda et al.(2004)]{matsuda2004} 
Matsuda, Y., Yamada, T., Hayashino, T., et al.\ 2004, \aj, 128, 569 

\bibitem[Matsuda et al.(2015)]{matsuda2015} 
Matsuda, Y., Nagao, T., Iono, D., et al.\ 2015, MNRAS, 451, 1141 

\bibitem[McMullin et al.(2007)]{mcmullin2007} 
McMullin J.~P., Waters B., Schiebel D., Young W., Golap K., 2007, ASPC, 376, 127 

\bibitem[Miller et al.(2016)]{miller2016} 
Miller, T.~B., Chapman, S.~C., Hayward, C.~C., et al.\ 2016, arXiv:1611.08552 

\bibitem[Nakamura et al.(2011)]{nakamura2011} 
Nakamura, E., Inoue, A.~K., Hayashino, T., et al.\ 2011, \mnras, 412, 2579

\bibitem[Obreschkow et al.(2009)]{obreschkow2009}
 Obreschkow, D., Croton, D., De Lucia, G., Khochfar, S., \& Rawlings, S.\ 2009, \apj, 698, 1467 

\bibitem[Oesch et al.(2014)]{oesch2014} 
Oesch, P.~A., Bouwens, R.~J., Illingworth, G.~D., et al.\ 2014, \apj, 786, 108 

\bibitem[Omont et al.(2013)]{omont2013} 
Omont, A., Yang, C., Cox, P., et al.\ 2013, \aap, 551, A115

\bibitem[Ono et al.(2014)]{ono2014} 
Ono, Y., Ouchi, M., Kurono, Y., \& Momose, R.\ 2014, \apj, 795, 5 

\bibitem[Orsi et al.(2014)]{orsi2014} 
Orsi, {\'A}., Padilla, N., Groves, B., et al.\ 2014, \mnras, 443, 799 

\bibitem[Pentericci et al.(2016)]{pentericci2016} 
Pentericci, L., Carniani, S., Castellano, M., et al.\ 2016, \apjl, 829, L11

\bibitem[Polletta et al.(2007)]{polletta2007} 
Polletta, M., Tajer, M., Maraschi, L., et al.\ 2007, \apj, 663, 81 

\bibitem[Popping et al.(2016)]{popping2016} 
Popping, G., van Kampen, E., Decarli, R., et al.\ 2016, \mnras, 461, 93 

\bibitem[Rangwala et al.(2011)]{rangwala2011} 
Rangwala, N., Maloney, P.~R., Glenn, J., et al.\ 2011, \apj, 743, 94 

\bibitem[Salpeter(1955)]{salpeter1955} 
Salpeter, E.~E.\ 1955, \apj, 121, 161

\bibitem[Sargsyan et al.(2012)]{sargsyan2012} 
Sargsyan L., et al., 2012, ApJ, 755, 171 

\bibitem[Seaquist et al.(2004)]{seaquist2004} 
Seaquist, E., Yao, L., Dunne, L., \& Cameron, H.\ 2004, MNRAS, 349, 1428 

\bibitem[Simpson et al.(2014)]{simpson2014} 
Simpson, J.~M., Swinbank, A.~M., Smail, I., et al.\ 2014, \apj, 788, 125

\bibitem[Smail et al.(2011)]{smail2011} 
Smail I., Swinbank A.~M., Ivison R.~J., Ibar E., 2011, MNRAS, 414, L95 

\bibitem[Smit et al.(2012)]{smit2012} 
Smit, R., Bouwens, R.~J., Franx, M., et al.\ 2012, \apj, 756, 14 

\bibitem[Speagle et al.(2014)]{speagle2014} 
Speagle, J.~S., Steinhardt, C.~L., Capak, P.~L., \& Silverman, J.~D.\ 2014, \apjs, 214, 15 

\bibitem[Suzuki et al.(2008)]{suzuki2008} 
Suzuki, R., Tokoku, C., Ichikawa, T., et al.\ 2008, \pasj, 60, 1347 

\bibitem[Swinbank et al.(2012)]{swinbank2012} 
Swinbank A.~M., et al., 2012, MNRAS, 427, 1066 

\bibitem[Swinbank et al.(2014)]{swinbank2014} 
Swinbank, A.~M., Simpson, J.~M., Smail, I., et al.\ 2014, \mnras, 438, 1267 

\bibitem[Stacey et al.(2010)]{stacey2010} 
Stacey, G.~J., Hailey-Dunsheath, S., Ferkinhoff, C., et al.\ 2010, \apj, 724, 957 

\bibitem[Takeuchi et al.(2003)]{takeuchi2003} 
Takeuchi, T.~T., Hirashita, H., Ishii, T.~T., Hunt, L.~K., \& Ferrara, A.\ 2003, \mnras, 343, 839 

\bibitem[Takeuchi et al.(2005)]{takeuchi2005} 
Takeuchi, T.~T., Buat, V., \& Burgarella, D.\ 2005, \aap, 440, L17 

\bibitem[Takeuchi et al.(2005)]{takeuchi2005b} 
Takeuchi, T.~T., Ishii, T.~T., Nozawa, T., Kozasa, T., \& Hirashita, H.\ 2005, \mnras, 362, 592 

\bibitem[Tamura et al.(2014)]{tamura2014} 
Tamura, Y., Saito, T., Tsuru, T.~G., et al.\ 2014, \apjl, 781, L39 

\bibitem[Uchimoto et al.(2012)]{uchimoto2012} 
Uchimoto, Y.~K., Yamada, T., Kajisawa, M., et al.\ 2012, \apj, 750, 116

\bibitem[Umehata et al.(2015)]{umehata2015} 
Umehata, H., Tamura, Y., Kohno, K., et al.\ 2015, \apjl, 815, L8 

\bibitem[Umehata et al.(2017)]{umehata2017} 
Umehata, H., Tamura, Y., Kohno, K., et al.\ 2017, \apj, 835, 98

\bibitem[Vallini et al.(2016)]{vallini2016} 
Vallini, L., Gruppioni, C., Pozzi, F., Vignali, C., \& Zamorani, G.\ 2016, \mnras, 456, L40 

\bibitem[Venemans et al.(2012)]{venemans2012} 
Venemans B.~P., et al., 2012, ApJ, 751, L25 

\bibitem[Wagg et al.(2010)]{wagg2010} 
Wagg J., Carilli C.~L., Wilner D.~J., Cox P., De Breuck C., Menten K., Riechers D.~A., Walter F., 
2010, A\&A, 519, L1 

\bibitem[Wagg et al.(2012)]{wagg2012} 
Wagg J., et al., 2012, ApJ, 752, L30 

\bibitem[Webb et al.(2009)]{webb2009} 
Webb, T.~M.~A., Yamada, T., Huang, J.-S., et al.\ 2009, \apj, 692, 1561 

\bibitem[Whiting(2012)]{whiting2012} 
Whiting, M.~T.\ 2012, \mnras, 421, 3242 

\bibitem[Williams, de Geus, \& Blitz(1994)]{williams1994} 
Williams J.~P., de Geus E.~J., Blitz L., 1994, ApJ, 428, 693 

\bibitem[Willott, Omont, \& Bergeron(2013)]{willott2013} 
Willott C.~J., Omont A., Bergeron J., 2013, ApJ, 770, 13 

\bibitem[Willott et al.(2015)]{willot2015} 
Willott, C.~J., Carilli, C.~L., Wagg, J., \& Wang, R.\ 2015, \apj, 807, 180 

\bibitem[Wolfire et al.(1995)]{wolfire1995} 
Wolfire, M.~G., Hollenbach, D., McKee, C.~F., Tielens, A.~G.~G.~M., \& 
Bakes, E.~L.~O.\ 1995, ApJ, 443, 152 

\bibitem[Wolfire et al.(2010)]{wolfire2010} 
Wolfire, M.~G., Hollenbach, D., \& McKee, C.~F.\ 2010, \apj, 716, 1191 

\bibitem[Vio \& Andreani(2016)]{vio2016} 
Vio, R., \& Andreani, P.\ 2016, \aap, 589, A20 

\bibitem[Yao et al.(2003)]{yao2003} 
Yao, L., Seaquist, E.~R., Kuno, N., \& Dunne, L.\ 2003, \apj, 588, 771 

\bibitem[Zhao et al.(2016)]{zhao2016} 
Zhao, Y., Lu, N., Xu, C.~K., et al.\ 2016, \apj, 819, 69 

\bibitem[Zhou et al.(2016)]{zhou2016} 
Zhou, L., Shi, Y., Diaz-Santos, T., et al.\ 2016, \mnras, 458, 772 

\end{thebibliography}
\end{document}